\documentclass[12pt,preprint]{aastex}

\begin{document}
\title{The Sgr~B2 X-ray Echo of the Galactic Center Supernova
  Explosion that \\ Produced Sgr~A East}

\author{Christopher L. Fryer\altaffilmark{1,2}, Gabriel
  Rockefeller\altaffilmark{1,2}, Aimee Hungerford\altaffilmark{3}, and
  Fulvio Melia\altaffilmark{1,4,5}}

\altaffiltext{1}{Department of Physics, The University of Arizona,
  Tucson, AZ 85721}
\altaffiltext{2}{Theoretical Division, LANL, Los Alamos, NM 87545}
\altaffiltext{3}{Computer and Computational Science Division, LANL,
  Los Alamos, NM 87545}
\altaffiltext{4}{Steward Observatory, The University of Arizona,
  Tucson, AZ 85721}
\altaffiltext{5}{Sir Thomas Lyle Fellow and Miegunyah Fellow.}

\begin{abstract}
  The possible impact Sgr~A East is having on the Galactic center has
  fueled speculation concerning its age and the energetics of the
  supernova explosion that produced it.  Using a combination of 1D and
  3D hydrodynamic simulations, we have carried out the first in-depth
  analysis of the remnant's evolution and its various interactions:
  with the stellar winds flowing out from the inner $\sim 2$ pc, with
  the supermassive black hole, Sgr~A*, and with the 50 km s$^{-1}$
  molecular cloud behind and to the East of the nucleus. We have found
  that, unlike previous estimates, a rather ``standard" supernova
  explosion with energy $\sim 1.5\times 10^{51}$ ergs would have been
  sufficient to create the remnant we see today, and that the latter
  is probably only $\sim 1,700$ years old. The X-ray Ridge between
  $\sim 9\arcsec$ and $15\arcsec$ to the NE of Sgr~A* appears to be
  the product of the current interaction between the remaining
  supernova ejecta and the outflowing winds.  Here again, the
  morphology and X-ray luminosity of this feature argue for a remnant
  younger than $\sim 2,000$ years. Perhaps surprisingly, we have also
  found that the passage of the remnant across the black hole would
  have enhanced the accretion rate onto the central object by less
  than a factor 2.  Such a small increase cannot explain the current
  Fe fluorescence observed from the molecular cloud Sgr~B2; this
  fluorescence would have required an increase in Sgr~A*'s luminosity
  by 6 orders of magnitude several hundred years ago.  Instead, we
  have uncovered what appears to be a more plausible scenario for this
  transient irradiation---the interaction between the expanding
  remnant and the 50 km s$^{-1}$ molecular cloud.  The first impact
  would have occurred about $1,200$ years after the explosion,
  producing a $2-200$ keV luminosity of $\sim 10^{39}$ ergs s$^{-1}$.
  During the intervening 300-400 years, the dissipation of kinetic
  energy subsided considerably, leading to the much lower luminosity
  ($\sim 10^{36}$ ergs s$^{-1}$ at $2-10$ keV) we see today.
\end{abstract}


\section{Introduction}

The conditions in the inner 3\,pc of the Galaxy are set in large part
by the complex interaction of over a dozen strong Wolf-Rayet star
winds and the enveloping $50$~km~s$^{-1}$ giant molecular cloud
(M-0.02-0.07), combined with the strong gravitational pull of the
central supermassive black hole (Sgr~A*) (see Melia \& Falcke 2001 for
a recent review).  The large number of massive stars in this compact
region suggests an additional influence---that of supernova
explosions.  Sgr~A East is the remnant of such an event.  An earlier
consideration of its interaction with the molecular cloud, based on
the energetics and time required to carve out the central cavity now
occupied by Sgr~A East \citep{Me89}, pointed to an unusually powerful
explosion, with an energy of $\sim 10^{52}$ ergs or greater, and a
remnant age exceeding $\sim 10,000$ years.  These estimates, however,
ignored the importance of the stellar winds in clearing out the medium
into which the supernova ejecta expanded following the incipient
event.  \citet{Roc05} have shown with detailed 3D hydrodynamic
simulations that, when one takes this additional factor into account,
the current morphology of Sgr~A East is actually consistent with a
normal supernova explosion energy ($\sim 10^{51}$ ergs).  In addition,
the relatively low gas density in the wind-filled region and the
consequent more rapid expansion of the remnant into the surrounding
medium leads to an inferred age much younger than $10,000$ years,
probably $< 2,000$ years.

\citet{Roc05} focused on the X-ray ridge to the NE of Sgr~A*, formed
by the interaction of the stellar winds---which emanate from within
the cavity enclosed by the circumnuclear disk (CND)---and the slowing
supernova ejecta expanding away from the site of the explosion that
produced Sgr~A East. In this paper, we discuss several additional
features that have emerged from our simulations, along with other
important implications of a young supernova remnant.  These include
the remnant's evolution, its impact on the supermassive black hole's
accretion, and the spatial distribution of heavy elements formed in
the supernova progenitor.  In particular, we wish to examine the
consequences of a young supernova remnant colliding with the
50~km~s$^{-1}$ cloud, focusing on the X-ray illumination this would
have produced on extended objects, such as the molecular cloud Sgr~B2,
$\sim 300$ light years to the NE of Sgr~A*. Sgr~B2's current
emission of a strongly-fluoresced Fe line appears to be the X-ray
``echo'' of that interaction, providing the best (circumstantial)
evidence of the impact the supernova would have had on the Galactic
center in the past several hundred years.

We begin by describing the current observational status of Sgr~B2 (\S\ 
2) and then take a step back to discuss the basic principles of shock
expansion in the Galactic center (\S\ 3).  Using a combination of
3-dimensional and 1-dimensional simulations, we then show how the
supernova shock would have affected the accretion rate onto Sgr~A*
(\S\ 4) and the concurrent evolution of the supernova remnant Sgr~A
East (\S\ 5), focusing on how these interactions might have accounted
for the irradiation of Sgr~B2.  We also present abundance distributions
produced by the explosion (\S\ 6) which, if measured, could provide us
with much tighter constraints on the supernova explosion energy.
Finally, we return to Sgr~B2 and conclude with a possible explanation
for the origin of the illumination that produced the current
fluorescent Fe emission.

\section{The Puzzle of Sgr~B2}

The region within $\sim 100$ pc of Sgr~A* contains giant molecular
clouds with a mean number density $\sim 10^4$ cm$^{-3}$ and a gas
temperature on the order of $60$ K \citep{Lis94}.  Over the past
decade, several instruments, including ASCA \citep{Koy96,M00} and {\it
  Beppo}SAX \citep{Sid99}, have revealed a source of bright
fluorescent Fe K$\alpha$ line radiation within the cloud Sgr~B2;
several other smaller clouds also exhibit strong $6.4$ keV line
emission, though with low absolute fluxes compared to Sgr~B2.  The
latter has a radius $\sim 10$--$20$ pc and a total enclosed mass $\sim
2$--$6 \times 10^6\;M_{\odot}$ \citep{Lis94}.  All of these
fluorescing clouds produce a $6.4$ keV line with an unusually large
equivalent width (EW $>$1--$2$ keV), though Sgr~B2 stands out with the
largest width, at $\approx 2$--$3$ keV. The surrounding continuum is
quite flat, and shows strong absorption below $4.5$~keV and a sharp Fe
K$\alpha$ absorption feature at $7.1$~keV.

The large EW is a strong indicator of how this fluorescent emission is
produced \citep{Sun93,Sun98,From01}.  A cloud radiates via X-ray
fluorescence when it is illuminated, either internally or externally,
by an X-ray source.  However, a steady source embedded within the
cloud produces an upper limit to the EW of only $\sim 1$ keV
\cite[see, e.g.,][]{Fab77,Vain80,From01}, regardless of how one
chooses the parameters.  The smaller molecular clouds might therefore
be marginally consistent with an internal illuminator \citep[see,
e.g.,][]{From01}.  However, Sgr~B2 must necessarily be illuminated
either by a time-dependent internal source whose flux has diminished,
or by an external source.  In the former case, the continuum will have
faded away relative to the line intensity; in the latter, we are not
directly observing the full ionizing flux. In both cases, the
equivalent width would be larger than in a situation where the
continuum spectrum of the irradiating source is still visible.

Recently, \citet{Rev04} have reported an association of the hard X-ray
source IGR J17475-2822 with Sgr~B2, showing that the ASCA ($3$--$10$
keV) and INTEGRAL/IBIS ($20$--$400$ keV) spectral components match
very well.  They showed that the combined spectrum at $3$--$200$ keV
can be well fit by a model in which X-rays from an external source,
possibly at the location of Sgr~A*, are scattered and reprocessed by a
homogeneous spherical cloud of cold, molecular hydrogen and helium
gas, with iron abundance $\sim 1.9$ times solar.

The possible identification of Sgr~A* as the external illuminator of
Sgr~B2 would provide some measure of its recent variability at X-ray
energies.  This association is motivated in part by the fact that the
iron emission in Sgr~B2 is strongest on the side of the cloud facing
Sgr~A* \citep{Koy96}.  We may be witnessing the X-ray echo, delayed by
300--400 years relative to the direct signal from the black hole, due
to the light travel time from the Galactic center out to Sgr~B2's
position. In this scenario, the fluorescent Fe emission would then be
direct evidence of the black hole's enhanced X-ray emissivity some
$300$ years ago.

Alternative scenarios seem to be falling out of favor \citep[see,
e.g.,][]{Rev04}.  For example, although the time-dependent internal
irradiator model can match the Fe line shape as well as the external
irradiator \citep{From01}, the lack of any significant variation in
the line flux with the passage of time argues against this geometry.
The large EW of the $6.4$ keV line implies that the primary source
should have faded away before the ASCA observation of 1993.  But the
light crossing time of the Sgr~B2 cloud is $\sim 30$--$60$ years, so
one might have expected to see a detectable decline of the $6.4$ keV
line flux in the 7 years between the ASCA and {\it Beppo}SAX
observations, unless the irradiation of Sgr~B2 is still ongoing
because not all the X-ray waves have yet reached the cloud.

Though it is very tempting to invoke Sgr~A* as the external
illuminator, there are several reasons for taking a cautious view of
this picture. Chief among them is the fact that this scenario would
require a change in the black hole's $2-10$ keV X-ray luminosity by a
factor of $\sim 10^6$ in only $300$ years, from $L_x \approx 5\times
10^{38}$ erg s$^{-1}$ \citep{Rev04} to the currently observed value of
$L_x \lesssim 10^{33}$ erg s$^{-1}$ \citep{Bag03}.  One of the
principal goals of this paper is to examine the role played by the
recent Galactic center supernova in illuminating Sgr~B2, which may
alleviate the difficulties described above.

\section{Supernova Shocks at the Galactic Center}

\begin{figure}
\plotone{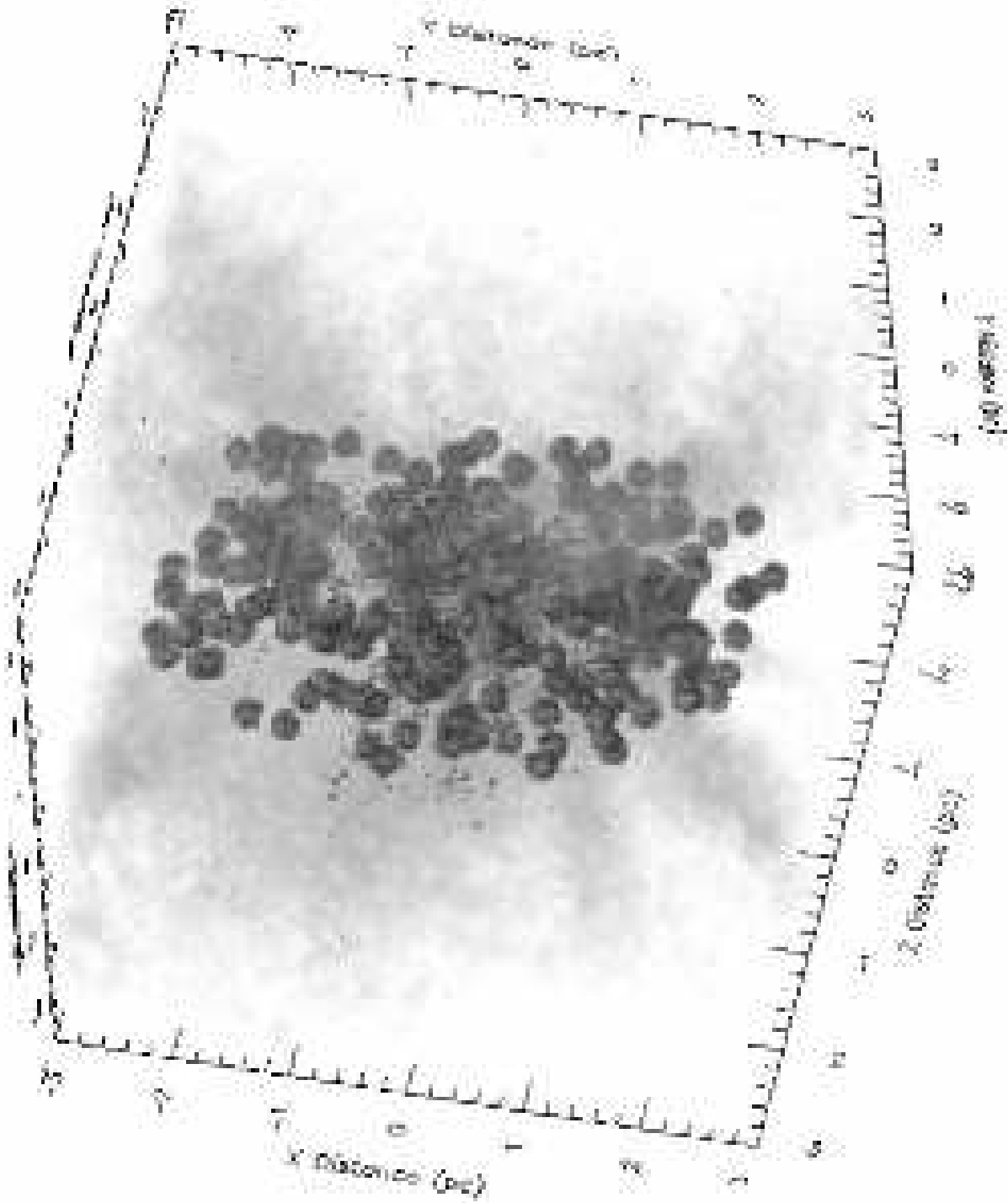}
\caption{Density profile of the Galactic Center.  The shading
  corresponds to number density values of ${\rm log}\ n \lesssim
  1$~cm$^{-3}$.  The contour corresponds to density values of
  ${\rm log}\ n > 1.5$~cm$^{-3}$.  The contour shows our low-filling
  factor disk along with the high-density central region around
  Sgr~A*.  The shading shows the very aspherical structure of the low
  density matter.}
\label{fig:3ddens}
\end{figure}

\begin{figure}
\plotone{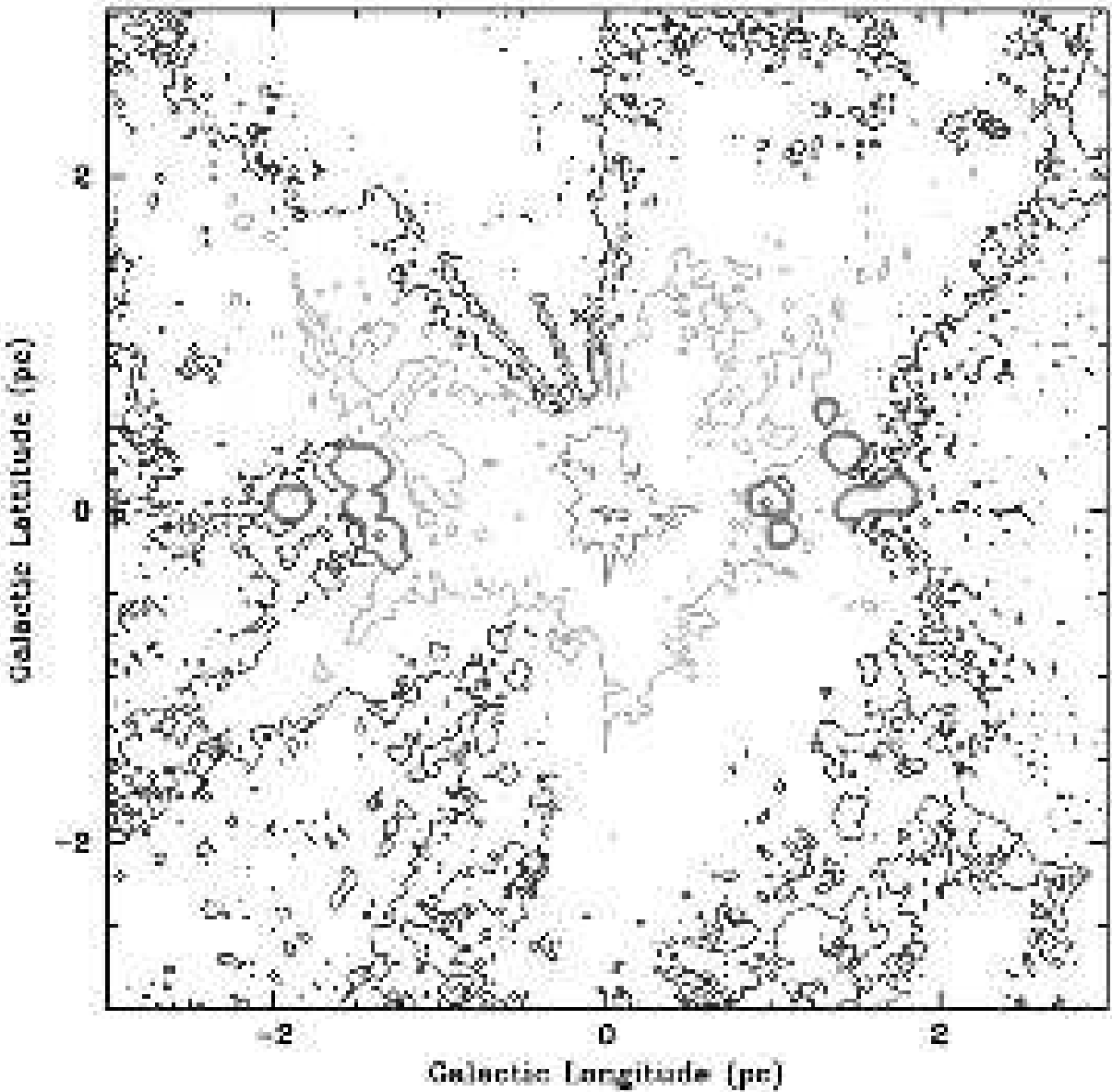}
\caption{Density contours of a 2-dimensional slice of the inner 3\,pc
  surrounding Sagittarius A*, 1 year after the launch of the supernova
  explosion at longitude = $-0.89$\,pc, latitude = $-1.47$\,pc.  The
  contours take mean densities from a 0.2\,pc slice centered on the
  supermassive black hole and correspond to densities of 1 (blue), 10
  (cyan), 100 (green), 1,000 (red), and $10^4$ (magenta) particles
  cm$^{-3}$.  The supermassive black hole is at 0,0.  The origin of
  the supernova may be seen as a small point density spike.  The
  circumnuclear disk is modeled as a number of dense spherical clumps
  with a low covering factor (magenta contours).  Note that the
  density in this inner Galactic center region is much lower than that
  of a typical molecular cloud and is highly asymmetric.}
\label{fig:2ddens}
\end{figure}

Molecular cloud remnants, inflowing plumes, and hot ionized bubbles
all combine to form a complex density structure across the Galactic
center, rendering it inappropriate for the ``spherically symmetric
cow'' approach preferred by most theorists.  These deviations from isotropy
are well-reflected in the anisotropic propagation of the supernova
shock.  Figure~\ref{fig:3ddens} shows the initial 3-dimensional
density structure used in our simulations.  We have here made the same
model assumptions as \citet{Roc05}, in which the density profile is
determined solely by the mass-losing stars interacting with the dense
circumnuclear disk (CND) within the gravitational potential of the
central supermassive black hole.  We have also assumed that the stellar 
wind ejecta have not changed significantly in the past $\sim
5,000$--$10,000$ years.  Figure~\ref{fig:2ddens} shows a 0.2\,pc slice
of this density structure centered on Sgr~A* at the time of the
explosion.

Before we discuss the propagation of the shock through this complex
density distribution, we review the spherically symmetric picture and
consider the 3 basic evolutionary phases of supernova remnants and
their shocks \citep{Cox72,Che74}:
\begin{itemize}
\item{Phase I (free-streaming):} The supernova explosion initially
  propagates essentially unimpeded by the surrounding medium.  This
  phase ends roughly when the supernova has swept up a mass equal to
  the pre-explosion mass of the progenitor.
\item{Phase II (adiabatic):} The remnant evolves into a second phase
  where cooling is still not important.  The shock can be described
  using adiabatic, self-similar blast wave solutions
  \citep{Sed59,Tay50}.
\item{Phase III (snow-plow):} The final phase occurs when radiative
  cooling becomes important.  In this phase, the thermal energy of the
  shock is rapidly radiated and the shock moves forward by momentum
  conservation alone.  This phase ends when the velocity of the shock
  decreases below the sound speed of the surrounding medium.
\end{itemize}

For a 15$\,M_\odot$ star, the end of the free-streaming phase ($t_{\rm
  free-streaming}$), and the remnant's radial extent ($R_{\rm
  free-streaming}$) at that time, are given, respectively, by the
expressions
\begin{equation}
t_{\rm free-streaming} \approx 2,000\, E_{51}^{-1/2} n^{-1/3} \, {\rm years}
\end{equation}
and
\begin{equation}
R_{\rm free-streaming} \approx 5.3\, n^{-1/3} \, {\rm pc,}
\end{equation}
as functions of supernova explosion energy $E_{51}$ (in units of
$10^{51}$\,ergs) and density of the surrounding medium $n$ (in units
of cm$^{-3}$).  Here we have assumed that the velocity is $\sqrt{2
  E_{\rm SN}/M_{\rm SN}}$ (where $E_{\rm SN}$ and $M_{\rm SN}$ are the
supernova energy and mass, respectively) which, however,
underestimates the lead shock speed.  We can use these spherical
estimates, combined with the density structure at the Galactic center
(Figure~\ref{fig:2ddens}), to follow the free-streaming phase along
specific paths.  That portion of the supernova shock that moves away
from Sgr~A* (where the number density is low: $\sim 1$ cm$^{-3}$) does
not decelerate significantly until it hits the 50 km s$^{-1}$
molecular cloud that surrounds the Galactic center (roughly 4\,pc from
the launch site of the supernova).  But the ejecta moving toward
Sgr~A* propagate through an increasingly dense medium.  The remnant on
this side of the explosion leaves the free-streaming phase more than
0.5\,pc away from Sgr~A*.

Beyond the free-streaming phase, but before radiative cooling becomes
important, the shock propagates adiabatically.  This phase lasts for a
period set by the cooling time of the shock.  \citet{Whe80} estimated
the radial and temporal extent of the shock for different cooling
functions.  When lines dominate, they find
\begin{equation}
t_{\rm cooling} \approx 110\, E_{51}^{0.22} n_4^{-0.56} \, {\rm years}
\end{equation}
and
\begin{equation}
R_{\rm cooling} \approx 0.29\, E_{51}^{0.29} n_4^{-0.43} \, {\rm pc,}
\end{equation}
where $n_4$ is the number density in units of $10^4$ cm$^{-3}$.
For the segment of the shock directed toward the Galactic center,
where densities are in the range of $10^3-10^4$ cm$^{-3}$, the shock
travels less than $\sim 0.5$ pc before cooling takes over.  From these
rough calculations, we might therefore expect the shock to just reach
Sgr~A*.  However, as we shall see in \S\ 4, the fact that the shock
can flow around this dense region, made impenetrable by the persistent
outward ram pressure of the stellar winds, means that the shock
actually never reaches the Galactic center. Correspondingly, the
ejecta moving away from the Galactic center have a much more extended
adiabatic phase; it lasts until they hit the molecular cloud, at which
point the phase ends almost immediately.

Timing is also important.  The total travel time for the ejecta to
reach the nucleus is just 600 years.  If the molecular cloud is 4 pc
away from the explosion site, our rough velocity estimate leads to the
shock leaving the adiabatic phase at roughly 1600 years.  Let us now
compare these results to the actual numerical calculations.

\section{Time-dependent Accretion Onto Sgr~A*}

We use both 1-dimensional and fully 3-dimensional simulations to trace
the evolution of the supernova remnant and to provide us with a basic
understanding of its effect on the Galactic center.  Of course, in the
1-dimensional case, we model the properties of the medium into which
the shock front expands in an angle-averaged sense. Under the
assumption of spherical symmetry, we have found that the shock passes
through both its free-streaming and adiabatic phases prior to reaching
Sgr~A*, but that it is less clear whether or not the shock actually
reaches Sgr~A* before cooling and assimilating into its surroundings.

What these spherically-symmetric simulations do not include are
multi-dimensional geometric effects.  Just as an ocean wave flows
around a rocky promontory, the supernova shock will flow around the
dense stellar-wind-filled region surrounding Sgr~A*.  Using the SNSPH
code described in \citet{FRW05}, we have modeled the propagation of
the supernova through the inner 3\,pc region surrounding the Galactic
nucleus \citep[for details, see][]{Roc05}.  The density profile was
taken from \citet{Roc04} and the supernova was assumed to occur with
an energy of $1.5\times10^{51}$\,ergs, in a progenitor with a mass of
15\,$M_\odot$ \citep{Hun05a}.  We placed it at longitude = $-0.89$~pc
and latitude =$-1.47$~pc relative to Sgr~A* in Galactic coordinates,
or $2$~pc due east of Sgr~A* in right ascension (but at the same
radial distance from us).  We have also modeled a more energetic
explosion ($\sim 1.2\times10^{52}$\,erg) by artificially increasing
the velocity of the ejecta by a factor 3; Table~\ref{table:sims}
summarizes the properties of both simulations.

\begin{figure}
\epsscale{0.8}
\plotone{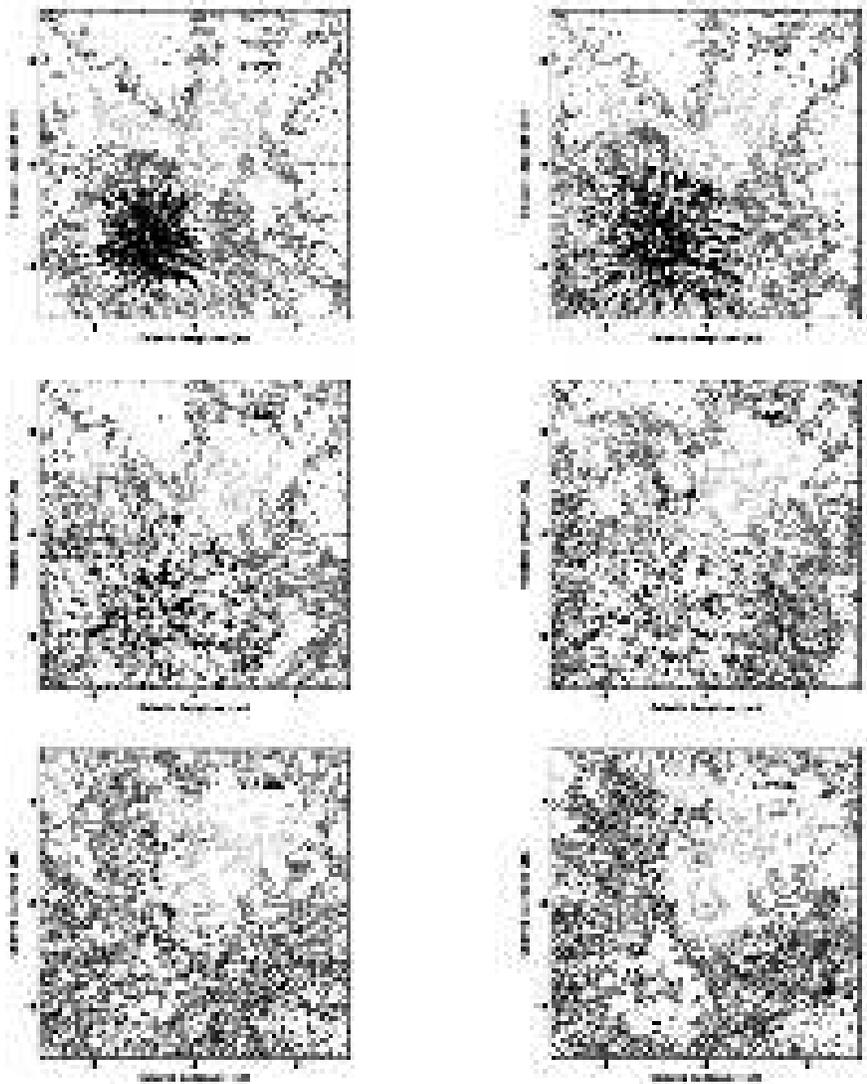}
\caption{Time series of a 0.2\,pc slice centered on Sgr~A* from our 
  simulation of a $1.5\times10^{51}$\,erg explosion.  Contours show
  density (same as Fig.~\ref{fig:2ddens}).  The vectors show the shock
  (black denotes supernova ejecta, red is the front of the shock
  composed of accelerated wind material.  Note that the shock flows
  around the dense central region surrounding Sgr~A* but does not
  penetrate closer than 0.4\,pc from the black hole.  The deepest
  penetration occurs at 650\,y.  At 1,400\,y the shock has cleared out
  most of the central 3 pc.  However, the region shadowed by the
  central 0.4 pc surrounding Sgr~A* is not cleared.  By 1,740 years
  the wind material has begun to reassert itself and is expanding back
  into the supernova ejecta.}
\label{fig:normaltime}
\end{figure}

Figure~\ref{fig:normaltime} shows a series of snapshots recording the
temporal evolution of the supernova explosion and the resulting
remnant from our 3-dimensional ($\sim10^{51}$\,erg) simulation.  The
contours indicate regions with different densities, while the vectors
highlight the supernova shock; the dark vectors correspond to the
supernova ejecta themselves and the light vectors indicate shocked
wind material.  The shock collides with and flows around the inner
0.4\,pc region primarily along paths of lowest density.  It ultimately
clears out much of the inner 3\,pc region, except for those portions
shadowed by the CND or by the outflowing winds from the central
0.4\,pc.  The depth to which our simulated shock penetrates in the
direction of Sgr~A* may be checked by simply comparing the ram
pressure of the supernova shock with that of the winds.  At 430 years,
the shock is within $\sim 0.4-0.5$ pc of the black hole.  The density
and velocity of the shock are roughly 3 particles cm$^{-3}$ and 4,000
km\,s$^{-1}$, respectively.  The corresponding values for the wind
outflow are 100 particles cm$^{-3}$ and 700 km\,s$^{-1}$.  The energy
density (or equivalently, the ram pressure, $\rho v^2$) of one flow is
equal to that of the opposing flow, and neither makes headway. This
roughly marks the time of maximum penetration of the supernova shock,
which eventually flows around the central region.  By 1,400 years
after the explosion, the ram pressure of the supernova shock decreases
below that of the wind, and the latter begins to reassert itself.

\begin{figure}
\epsscale{0.8}
\plotone{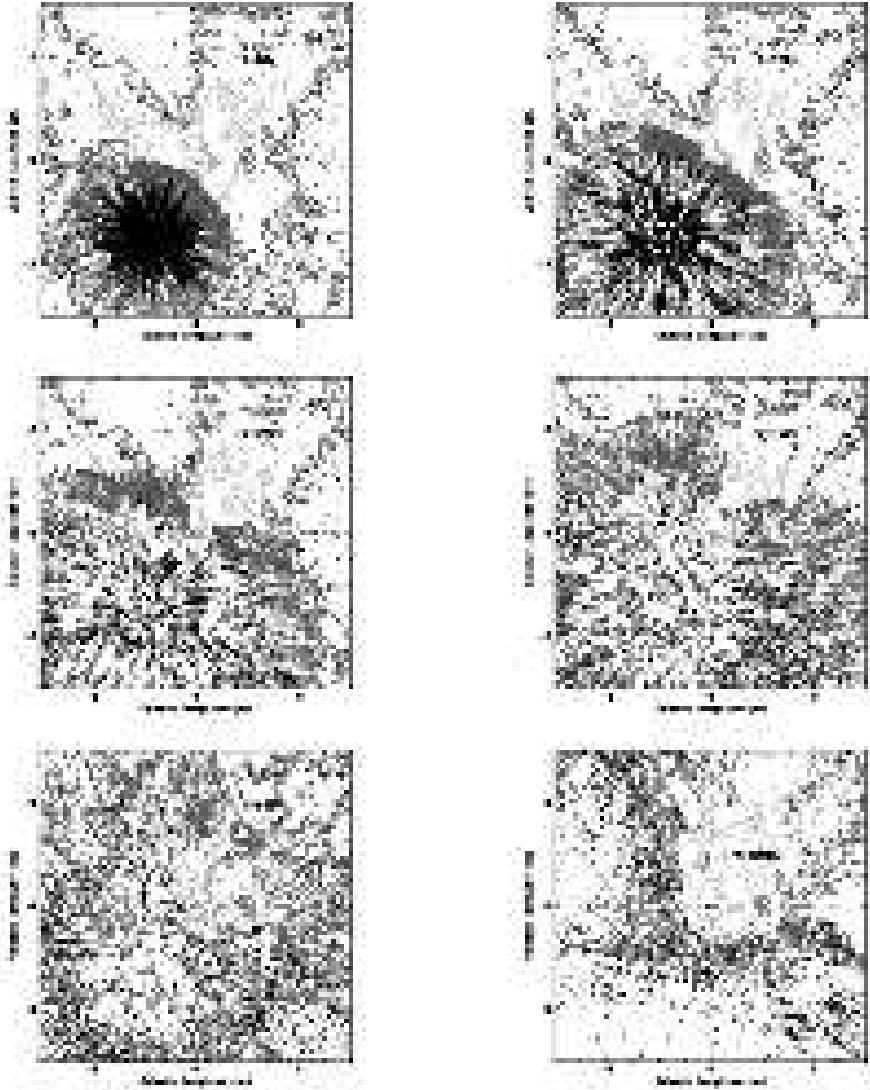}
\caption{Same as Fig.~\ref{fig:normaltime}, but now for the 
  $1.2\times10^{52}$\,erg explosion.  The shock penetrates to within
  0.1-0.2\,pc of Sgr~A*, reaching maximum penetration at only 200-270
  years.  It clears out the Galactic center more comprehensively than
  the $\sim 10^{51}$ erg explosion, but some material is still
  shadowed by the dense stellar wind region surrounding Sgr~A*.  By
  1,200 years, the stellar winds have begun to reassert themselves.}
\label{fig:energtime}
\end{figure}

In the $\sim 10^{52}$ erg simulation (shown in
Figure~\ref{fig:energtime} in a similar series of snapshots), the
shock moves much faster and penetrates deeper into the stellar wind
region surrounding Sgr~A*.  By 170 years, the shock is within 0.2\,pc
of the black hole, but as in the standard-energy simulation the shock
flows around the dense central region, and even at the distance of
closest approach---roughly 270 years after the explosion---the
supernova shock never gets closer than 0.1\,pc from Sgr~A*.  After
1,200 years, the shock has swept through the entire 3 pc central
region, clearing out most of the material except for the CND and the
stellar winds, which have begun to reassert themselves beyond 0.4 pc.

\begin{figure}
\plotone{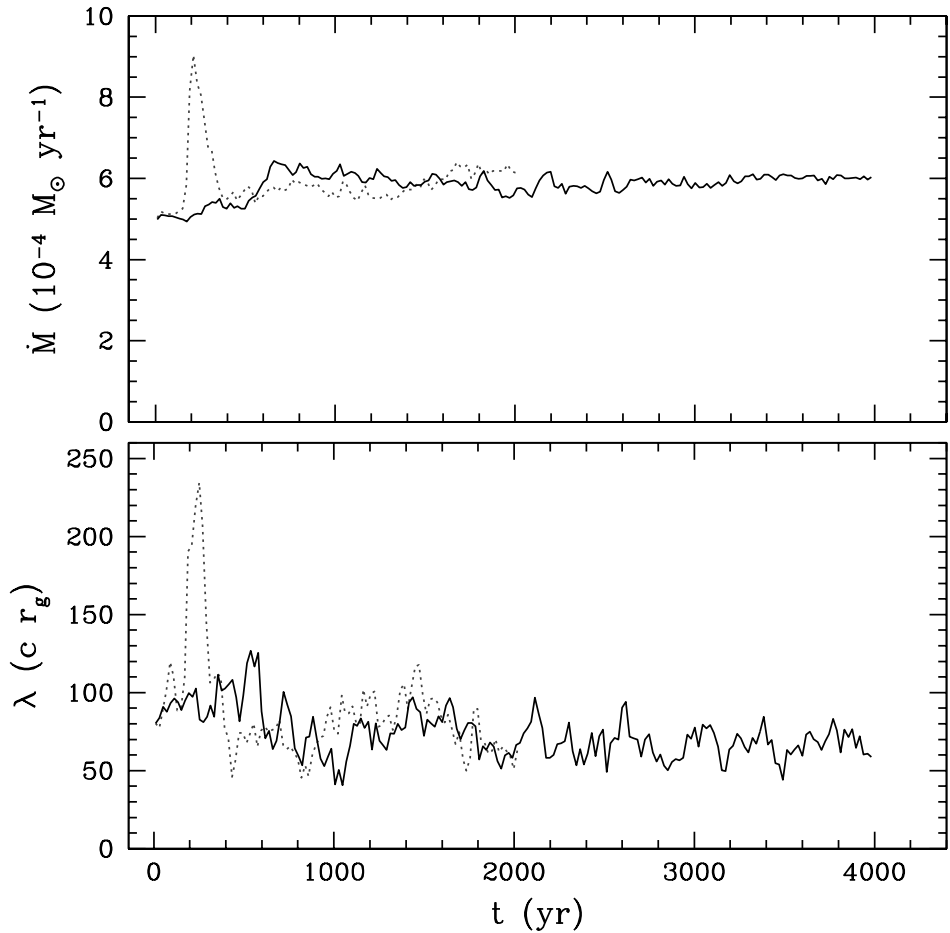}
\caption{The mass accretion rate (top) and accreted specific angular 
  momentum (bottom) as functions of time for the standard (solid line)
  and energetic (dotted line) simulations.  The standard simulation
  undergoes a small ($\sim 20$\%) change in accretion rate around the
  time of closest approach of the supernova shock to Sgr~A* ($\sim
  650$\,yr), while the energetic simulation shows an increase in
  $\dot{M}$ of nearly a factor $2$ for a short period of time.
  Similarly, the accreted specific angular momentum ($\lambda$, in
  units of $cr_g$, where $r_g\equiv GM/c^2$) changes by $\lesssim
  20$\% in the standard simulation, but undergoes a brief increase of
  more than a factor $2$ in the energetic simulation.}
\label{fig:mdot}
\end{figure}

Even though the shock doesn't actually reach Sgr~A*, it can still
affect the medium there, and possibly alter the rate at which matter
(and its angular momentum) is accreted onto the black hole.
Figure~\ref{fig:mdot} shows the accretion rate and accreted specific
angular momentum as a function of time for both simulations.  Even the
high-energy simulation exhibits an increase in accretion rate of less
than a factor 2 around the time of closest approach of the
supernova shock to the black hole; the standard-energy simulation shows
an increase of only $\sim 20$\%.  The accreted specific angular
momentum shows similar variation; the specific angular momentum
accreted in the high-energy simulation jumps by more than a factor 
2 for a short period, but the normal-energy simulation shows almost no
change at all.

In summary, our simulations demonstrate that even with a $\sim
10^{52}$\,erg explosion, a supernova (or ``hypernova'', as these rare
energetic explosions are often called) does not penetrate all the way
to Sgr~A* and does not significantly alter the accretion rate.  It is
difficult to imagine any scenario in which such minor changes in the
accretion rate can result in a brightening of Sgr~A* by a factor of a
million, allowing it to be the transient irradiator of Sgr~B2 several
hundred years ago.  If Sgr~B2 was indeed illuminated by emission at
the Galactic center, the source of those X-rays must lie beyond
Sgr~A*.

\section{Evolution of the Sgr~A East Remnant}

As the supernova shock moves outward, it ultimately hits the 50 km
s$^{-1}$ molecular cloud behind and to the east of Sgr~A*.  The last
vestige of this interaction is visible now as Sgr~A East. However, the
dense molecular cloud lies well outside the simulation space of our
3-dimensional calculations.  Currently, the inclusion of such a large
volume, with the spatial resolution we need in order to produce images
such as those shown in earlier figures, is beyond our present
computational capability in 3 dimensions.

In this section, we examine the evolution of Sgr~A East, focusing on
its viability to act as a source for the past illumination of Sgr~B2.
To do so, we employ a simplified version of the 1-dimensional
Lagrangian code developed by \citet{Fry99}.  We include in this code
an approximate cooling term, using mean values of the cooling function
given in \citet{Sut93}.  We vary the magnitude of this energy loss
rate and find, within the range of values given by our temperatures,
that the exact value of the cooling rate does not affect our results
considerably.  What is important, however, is the density profile of
the medium through which the shock is propagating.

In their detailed study of molecular gas in the central $10$ parsecs
of the Galaxy, \citet{Herr05} examined the interaction between the
Sgr~A East shell and the 50 km s$^{-1}$ cloud and concluded that the
expansion of the former apparently did not move a significant amount
of the latter's mass. This is consistent with the results of our
simulation, in which the supernova ejecta at first moved rather quickly 
through the medium surrounding Sgr~A*, which had been mostly cleared
out by the powerful winds of stars situated within $\sim 2-3$ parsecs
of the black hole.  But Sgr~A East is clearly interacting with the
50~km~s$^{-1}$ cloud now, as evidenced by the presence of seven 1720
MHz OH maser emission regions within several arcmins of the Galactic
center \citep{Yusef96,Yusef99}.  This transition of the OH molecule is
a powerful shock diagnostic and is collisionally pumped by H2
molecules at the site where C-type supernova shocks drive into
adjacent molecular clouds.  Most of these maser spots are located to
the SE of Sgr~A*, at the boundary of Sgr~A East and M-0.02-0.07.  In
addition, Zeeman splitting measurements suggest that the magnetic
field at these locations is of order $2-4$ mG.  Both the relatively
high intensity of this field, and the intense OH maser emission,
indicate that the shock at the interface between Sgr~A East and
M-0.02-0.07 must be very strong, since the impact is compressing the
gas and the field lines.

The present interaction region between Sgr~A East and M-0.02-0.07
appears to be $\sim 1'-1.5'$ in projection from Sgr~A*
\citep{Yusef96,Herr05}.  At the distance to the Galactic center, this
corresponds to $\approx 2.4-3.6$ parsecs; taking projection into
account, we infer $\sim 4$ parsecs as a reasonable estimate of the
distance between the interaction site and Sgr~A*.  Thus, with our
chosen supernova site 2 parsecs due east (in Right Ascension) of
Sgr~A*, it would have taken $\sim 1200$ years for the shock front to
reach the molecular cloud traveling at a speed of $v \sim \sqrt{2
  E_{\rm SN}/M_{\rm SN}} \sim 2,500$ km s$^{-1}$. 

\begin{figure}
\plotone{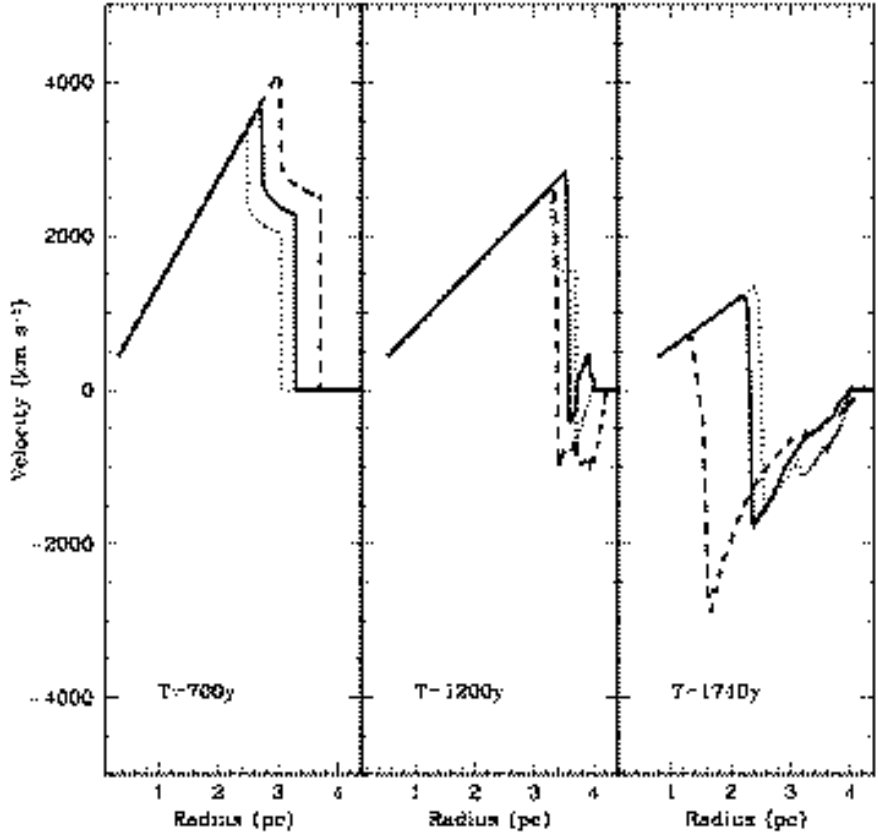}
\caption{Velocity versus radius at 3 snapshots in time from our
  1-dimensional simulation, modeling the encounter of the supernova
  shock with the surrounding 50 km s$^{-1}$ molecular cloud.  The
  solid line denotes our standard model, which is essentially a
  1-dimensional spherical version of the $1.5\times10^{51}$\,erg
  supernova explosion modeled in our 3-dimensional simulation.  We
  assume a 4\,pc windswept medium (with density $n=1\; {\rm cm^{-3}}$)
  surrounded by a dense ($n=10^4\; {\rm cm^{-3}}$) molecular cloud.
  The dotted line is the same explosion, but with a slightly higher
  density in the windswept medium ($n=2\; {\rm cm^{-3}}$) and a
  gradient in the molecular cloud density rising ultimately up to
  $n=10^5\; {\rm cm^{-3}}$.  The dashed line is an explosion with a
  20\% increase in energy, and a 4.2\,pc windswept region, but with
  the same densities as in our standard model.}
\label{fig:1dvel}
\end{figure}

Figure~\ref{fig:1dvel} shows several snapshots in time of the velocity
of propagation for three 1-dimensional explosion calculations.  The
basic setup of these models is a diffuse ($n < 10$ cm$^{-3}$)
wind-swept region with an outer dense ($n > 10^4$ cm$^{-3}$) molecular
cloud, roughly starting at 4\,pc.  As we would expect from the Sedov
blast wave similarity solution \citep{Sed59,Tay50}, the shock
decelerates as it propagates through the diffuse wind-swept medium.
If the diffuse density were higher, the shock would decelerate faster
and reach the molecular cloud at a later time.  If the explosion
energy were higher, the shock would move faster and hence reach the
molecular cloud earlier.  In all cases, the shock essentially hits a
wall at the molecular cloud and bounces back, sending a reverse shock
through the diffuse, lower density region.

The energy dissipated when the shock interacts with the molecular
cloud can be a significant, albeit transitory, source of high-energy
radiation.  Supernova remnants interacting with molecular clouds are
efficient electron accelerators and sources of hard X-ray and
$\gamma$-ray emission \citep{Byk00}.  The energy spectrum of the
nonthermal electrons is shaped by various processes, including first
and second-order Fermi acceleration in a turbulent plasma, and energy
losses due to Coulomb, bremsstrahlung, synchrotron, and inverse
Compton interactions.  The spectrum produced by these particles
between $\sim 1$ keV and $\sim 1$ MeV is essentially a power law,
$\nu\,F_\nu\propto\nu^{-\alpha}$, with $\alpha\sim 0.25$. The
efficiency of energy transfer from the shock flow to the nonthermal
electrons is roughly $5\%$, though the actual value depends on the
velocity of the shock, the density in the cloud, and the radiative
efficiency; the efficiency may be lower, but under some conditions, it
could be as much as a factor of 2 greater.  Since the detailed
calculation of the particle acceleration and radiation is beyond the
scope of the present paper, we will here simply adopt $5\%$ efficiency
as the fiducial value, and calculate the overall X-ray/$\gamma$-ray
luminosity from the Sgr~A East/M-0.02-0.07 interaction site by
estimating the shock energy dissipation rate from our 1-dimensional
simulation, and assuming that all of the nonthermal particle energy is
eventually radiated.  We note that with our simplified 1-dimensional
simulations, the reverse shock ultimately produces many regions of
compression, bouncing backwards and forwards. However, these
subsequent shocks are unlikely to be as strong as the first in the
aspherical geometry of the Galactic center.  Here, we focus only on
the energetics of the first (or leading) impact.

\begin{figure}
\plotone{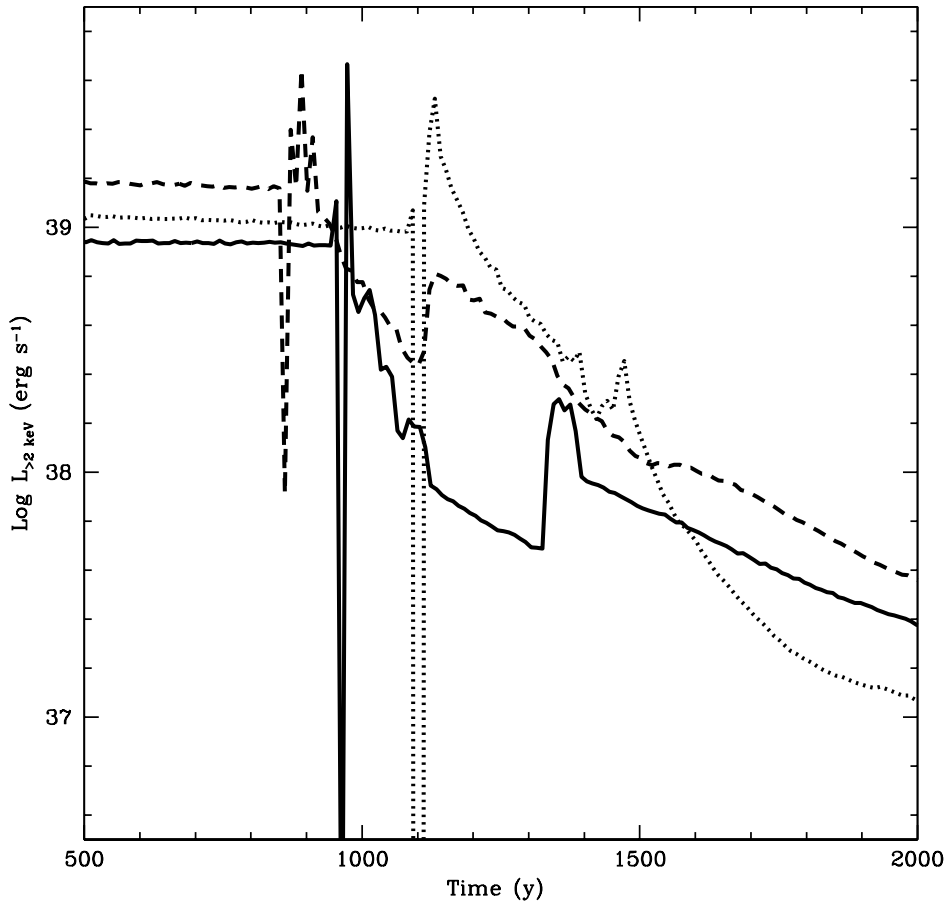}
\caption{The $2-200$ keV luminosity versus time for the 3 shock models
  shown in Figure~\ref{fig:1dvel}.  Our best-fit model, with a
  slightly increased density in the windswept region and a density
  gradient (dotted line), produces considerable high energy emission
  (enough to illuminate Sgr~B2) up to 400 years ago (assuming the
  supernova remnant is 1,700 years old), yet the high energy flux has
  decreased significantly by the current epoch.  Note that the $2-10$
  keV flux is $\sim 20\%$ of this total (i.e., bolometric) high energy
  value.  Geometric effects, ignored in these spherically symmetric
  simulations, will alter these results somewhat, but these
  calculations do establish the plausibility of a model in which Sgr~A
  East illuminated Sgr~B2 when it first encountered the giant
  molecular cloud.}
\label{fig:1dlum}
\end{figure}

The $\sim 2$--$200$ keV luminosity resulting from the interaction we
are simulating here is shown as a function of time in
Figure~\ref{fig:1dlum}.  This light curve is calculated with the
conservative assumption, described above, that only 5\% of the
dissipated energy in the leading shock is converted into photons above
$2$ keV. Since the 50 km s$^{-1}$ molecular cloud does not completely
envelop the Galactic center, we also assume that the interaction site
occupies only $4\pi/3$ of solid angle; this estimate is, of course,
only a rough approximation, but it does not significantly impact our
conclusion.

The result shown in Figure~\ref{fig:1dlum} clearly establishes the
possibility that the interaction between the supernova that created
Sgr~A East, and the giant molecular cloud, produced the transient
X-ray flux whose echo we see today in the Fe fluorescence of Sgr~B2.
This conclusion comes with several caveats, however, mostly having to
do with uncertainties in the overall irradiating luminosity: first,
recall that the shock will bounce off of the molecular cloud and send
a reverse shock back through the outflowing ejecta, causing the total
$>2$ keV emissivity to be higher than the value we have calculated
here; this is potentially good for the model. Second, when we allow
for additional dimensions in the calculation, the shock may flow
around the molecular cloud, so the shock in Sgr~A East may not be as
strong as we find in our 1-dimensional simulation; of course, this
will lower the yield of nonthermal particles, and hence the $2-200$
keV luminosity.  Which of these factors wins out may ultimately
determine whether or not this model is correct. Given the level of
sophistication of our current calculations, we can only say that both
the luminosity and the timing associated with the peak of the
dissipation seem to be those required to account for the properties of
Sgr~B2.  Our calculation shows that within the last $\sim 400-500$
years, over $10^{39}$ erg s$^{-1}$ were released in photons with
energy above 2 keV.

It is beyond the scope of the present paper to calculate in detail the
spectrum of the irradiating flux, but we note from the work of
\citet{Byk00} that the radiation produced by the nonthermal particles
is essentially a power law with flux $F(\nu) \propto \nu^{-0.75}$.
Thus, the integrated luminosity in the $2-10$ keV range should be
$\sim 20\%$ of the total.  With reference to {\it Chandra's} spectral
band, our predicted X-ray flux is therefore roughly 1/5th the value
shown in Figure~\ref{fig:1dlum}. Given that the peak irradiance
occurred $\sim 400$ years ago, the $2-10$ keV flux level now is
therefore consistent with {\it Chandra's} current measured X-ray
luminosity of $\sim 10^{36}$ ergs s$^{-1}$ from the Sgr~A
East/molecular cloud interaction region.  And since our calculated
light curve is also a good match to the required illumination of
Sgr~B2 by a $\sim 2-200$ keV spectrum with a peak luminosity of $\sim
10^{39}$ ergs s$^{-1}$ some $300-400$ years ago \citep{Rev04}, we see
that both the temporal variation of the high-energy flux, and its
associated spectrum, are consistent with all the currently available
data.

\section{Other Constraints on the Recent Galactic Center Supernova}

Two uncertainties dominate our solution of the Sgr~B2 illumination
problem: the supernova explosion itself and the environment through
which the explosion travelled.  Understanding the environment requires
first trying to get a full 3-dimensional structure from observations
and then extrapolating that structure backward in time (a process that
also requires knowledge of the supernova explosion and its progenitor 
star).  Here, we will instead focus on possible observations
that can help constrain the supernova explosion only.  In particular,
we would like to study the issue of the supernova remnant's age and
related issues concerning the origin of the explosion and its energy.

\citet{Roc05} found that the structure of the X-ray ridge and its
X-ray flux constrained the remnant's age.  They assumed a roughly
``standard" ($1.5 \times 10^{51}$\,erg) explosion, a
radial position of the origin of the supernova set to the radial
position of Sgr~A*, and a central region whose structure is dominated
by the interaction of the stellar winds with the CND and a central
supermassive black hole.  In this paper, we have expanded this study to
include a much more energetic $\sim 1.2\times10^{52}$\,erg explosion.
By carrying out a similar study of the X-ray luminosity and shape of the 
X-ray ridge, we estimate the age of the remnant produced by the
strong explosion to be 700\,y.  It may be difficult to explain the 
current luminosity of Sgr A East with such a young, strong supernova 
explosion.

\begin{figure}
\plotone{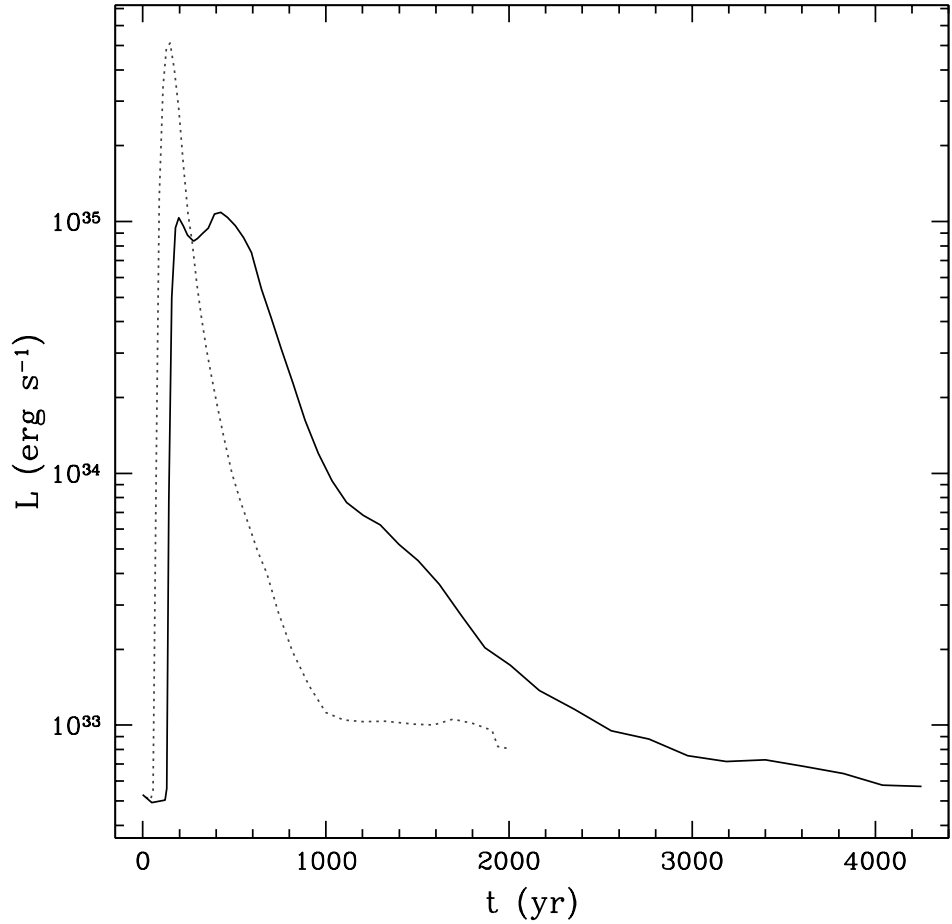}
\caption{The 2--10 keV X-ray luminosity from a $170^\circ$ arc on the 
  simulated sky between $9^{\prime\prime}$ and $15^{\prime\prime}$ to
  the east of Sgr~A* (toward the supernova).  The solid line shows the
  2--10 keV luminosity from the standard simulation; the dotted line
  shows the luminosity from the energetic simulation.}
\label{fig:3dlum}
\end{figure}

Clearly, the remnant's age corresponding to either explosion energy is
much less than previously thought.  But can we place an upper limit on
this age?  Figure~\ref{fig:3dlum} shows the 2-10\,keV luminosity of
the X-ray ridge.  Observations of this ridge can be fit with a
two-component model to its luminosity: a 1\,keV component with a flux
of $5.8\times10^{-13}$~erg~cm$^{-2}$~s$^{-1}$, and a 5.6\,keV
component with a flux of $3.92\times10^{-13}$~erg~cm$^{-2}$~s$^{-1}$;
these fluxes translate to luminosities of
$4.4\times10^{33}$~erg~s$^{-1}$ and $3.0\times10^{33}$~erg~s$^{-1}$,
respectively, assuming a distance of $8.0$~kpc to the Galactic center.
If we assume that only the 5.6\,keV component is actually associated
with the interaction that produces the ridge, we find that the
standard simulation matches this luminosity at $\sim 1,700$~y; the
energetic simulation matches it at $\sim 700$~y.  A factor 2
uncertainty in the flux would place an upper age limit at $\sim
2,100$~y for the standard explosion and at $\sim 900$~y for the
energetic event.

\begin{figure}
\plotone{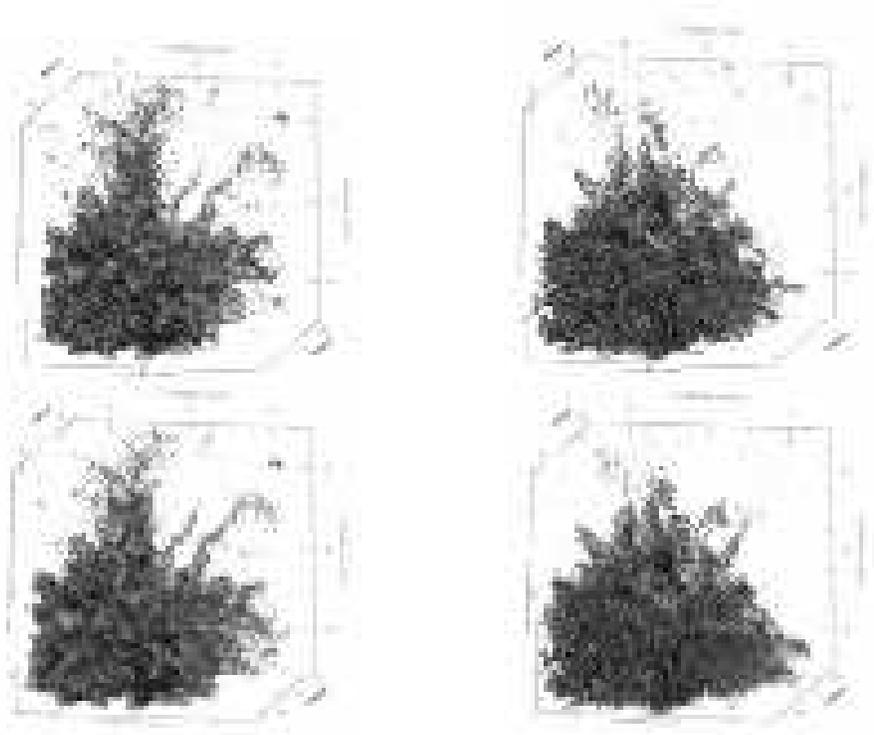}
\caption{Abundance plots for both the standard (left column) and
  energetic (right column) explosions.  The top panels show the nickel
  (iron) abundance surrounded by the silicon abundance.  Note that at
  this time, because the silicon was produced further out, it had a
  higher expansion velocity and has therefore mixed further out.  The
  silicon-to-iron ratio should be much higher than typical solar
  abundances at these larger distances.  The bottom panels show the
  inner titanium (made in the outer part of the region where the iron
  is made) surrounded by the magnesium abundance for comparison.}
\label{fig:3dabund}
\end{figure}

Such a short remnant age has serious implications on the chemical
enrichment from this supernova.  The $^{56}$Ni (which decays into the
main iron products of the supernova) is produced only in the inner
layers of the star.  The explosion quickly develops into a homologous
expansion (meaning that the ejecta velocity is proportional to the
radius); the iron, produced in the inner layers of the ejecta, is
moving slowly.  In our standard explosion, it is never moving faster
than about 2000\,km\,s$^{-1}$ (in contrast, the highest velocity in
the energetic explosion is roughly 6000\,km\,s$^{-1}$).  If the iron
moved at this high velocity without decelerating, our standard-energy
supernova would not enrich regions in the Galactic center beyond
3.6\,pc, but because the iron does decelerate (along with the rest of
the shock), this number is closer to 2\,pc (Figure~\ref{fig:3dabund}).
In the energetic explosion, the iron travels faster and penetrates
farther into the Galactic center medium.  Indeed, the intermediate
weight elements (e.g. silicon and magnesium) have traveled beyond our
computational grid boundaries.  From Figure~\ref{fig:3dabund}, we see
that elements made near the iron layer are, like the iron, limited to
a reduced region around the origin of the supernova; elements made
further out in the star extend much farther into the surrounding
region.  In addition, there is a zone shadowed by the stellar winds
around Sgr~A* that is not enriched at all by the supernova.

There are two major caveats to these abundance plots.  First, we have
assumed a spherically symmetric explosion and subsequent expansion.
But we now know that core-collapse supernovae are far from symmetric
\citep[see][for reviews]{Hun03,Hun05a}.  Asymmetries in the explosion
will allow some iron to mix much further out in the star, and due to
the homologous outflow, achieve higher velocities.  However, the bulk
of the iron will not reach velocities significantly different from
what we have obtained in our spherical models.  We also did not
incorporate into our energetic event the greater yield of heavy
elements in a stronger explosion \citep{Hun05b}.  The more energetic
event could produce a factor 2 more iron, and 10 times more $^{44}$Ti,
but the basic distribution would remain the same.

The X-ray ridge observations push for a young supernova remnant.
Using the X-ray ridge ages, we expect the iron to still be close to
the origin of the supernova explosion (further out for the strong
explosion than for the weak explosion).  If the remnant truly is
5,000-10,000\,y old, the iron should be well mixed all the way into
Sgr A*.  High resolution abundance maps could both confirm or
cast-into-doubt the young age predicted by the X-ray ridge.  In
addition, such maps could help locate the origin of the explosion.
Combined with better models of Sgr A East, such information will also
constrain the supernova energy.  One way to get this information is to
look for evidence of compositional variation in the dust grains and
molecular gas in the denser regions: e.g., in the CND, the Northern
Ridge, or the Western Streamer \citep{Herr05}.  Depending on the
supernova energy, age, and location, one or more of these dense
regions may be enriched by metals from the supernova itself.  For the
standard supernova explosion simulation, we expect only the CND to be
enriched by iron or even silicon.  Since elements like silicon and
magnesium are primarily tied up in dust grains, depending sensitively
on the physical conditions, accurate abundance measurements may be
difficult to obtain.

Emission from radioactive elements is less sensitive to the physical
conditions and could also provide some clues.  The major intermediate
age elements are $^{44}$Ti and $^{59}$Ni.  Unfortunately, for the
standard supernova explosion energy, the age of the remnant is
so much larger than the $\sim 60$\,y half-life of $^{44}$Ti, that the
resultant high energy flux from its decay is 10 million times fainter 
than that of 1987a (for the energetic explosion with its likely
higher $^{44}$Ti yield and younger age, this value is still a million
times fainter than 1987a).  This is well beyond any detection
limit.  However, the $^{59}$Ni flux (with its 75,000\,year lifetime)
could be as high as $2.5\times10^{-7} {\rm cm^{-2} \, s^{-1}}$
(possibly even higher for the energetic explosion).  Although beyond
detectable limits of current instruments \citep{Lei02}, this flux may
be observed with an improved generation of high-energy telescopes.

None of these observational constraints are easy to obtain.  We
provide this information to encourage and justify experimental and
observational programs that might be able to shed more light on this
complex problem.

\section{Conclusions}
The fact that Sgr~A East had little impact on Sgr~A*'s accretion rate
as the remnant passed across the Galactic center is of some
consequence to the question of how Sgr~B2 and other nearby molecular
clouds produce their strong fluorescent Fe line emission at $6.4$ keV.
This puzzle has been viewed as an important indicator of past
high-energy activity at the Galactic center, though with very little
guidance from gas dynamic studies---until now.

This is where our results from the 3-dimensional and 1-dimensional
simulations enter the discussion.  One might have thought that the
passage of the supernova remnant's front across Sgr~A* could have
triggered a significant increase in the rate of gas infall, possibly
even producing an enhanced accretion rate onto the black hole to fuel
its $\sim 10^6$ factor increase in X-ray luminosity.  The implication
of this model would be that the supernova shock passed through the
Galactic center some $300-400$ years ago.  As we have seen, however,
the gas dynamics within the inner $3-4$ parsecs of the Galaxy negate
this possibility, because the strong, cumulative outflux of matter
from the interior of the CND prevents Sgr~A East from penetrating
closer than $\sim 0.2$ parsecs from Sgr~A*. At best, the black hole's
accretion rate could have increased by possibly $20-30\%$, far short
of the value required to account for the irradiation of Sgr~B2.

Instead, our modeling of Sgr~A East's interaction with the Galactic
center has produced what we believe is a far more plausible scenario
for the variable illumination of Sgr~B2 several hundred years ago. As
we saw earlier, the remnant's shock front apparently reached the
Galactic center some 160 years after the supernova explosion, and
swept around the inner $0.2-0.3$ parsec region in $\sim 650$ years.
Our modeling of the X-ray ridge NE of Sgr~A* suggests that we are
viewing the interaction between the remnant and the winds flowing out
from the ionized cavity $\sim 1,700$ years after the supernova event.
There is strong observational evidence that Sgr~A East is now also
interacting with the so-called 50 km s$^{-1}$ molecular cloud behind
Sgr~A* \citep{Yusef99}. But this interaction should have produced an
intense X-ray/$\gamma$-ray glow when the impact first occurred, due to
the initial rapid dissipation of kinetic energy flux into heat,
nonthermal particle acceleration and radiation.  Given the proximity
of the supernova event site to the Galactic center (only several
parsecs away from Sgr~A*), the ensuing irradiation
of Sgr~B2 would still show the tell-tale characteristics of a Galactic
center source, albeit now a diffuse X-ray/$\gamma$-ray emitter, rather
than a point source associated directly with Sgr~A*.  The key
constraint is that Sgr~A East's X-ray/$\gamma$-ray glow should not
have ended more than $\sim 400$ years ago---approximately $1,350$
years after the supernova explosion---this being the light travel time
between the interaction site and Sgr~B2.

There may even be related evidence that the X-ray/$\gamma$-ray glow
may have persisted closer to the present time, perhaps to within the
past 100 or 200 hundred years. In their analysis of Sgr~A East and its
X-ray properties, \citet{Mae02} showed that the ionized gas halo into
which Sgr~A East is currently expanding (i.e., in regions other than
where it is colliding with the 50 km/s cloud) may have been ionized by
the same irradiator that produced the current fluorescence in Sgr~B2.
For an ambient gas density $\sim 10^3$ cm$^{-3}$, the required
luminosity would have been $\sim 10^{40}$ ergs s$^{-1}$, but since the
recombination time in such a gas is shorter than $\sim 300$ years, the
(still) high ionization fraction in the ISM argues for a period of
irradiation extending to well within 300 years of the present.
However, the current X-ray luminosity from the Sgr~A East shell,
including the region of interaction with the 50 km/s cloud, places a
rather severe constraint on how rapidly the X-ray/$\gamma$ ray glow
must have subsided to its present value. This $2-10$ keV limit is
$\approx 10^{36}$ ergs s$^{-1}$, if we take the whole shell into
account. A reduction in the solid angle subtended by the interaction
zone at the site of the explosion would lower this level even further.

The circumstantial evidence for a youthful Sgr~A East remnant is
building. We now have several observational indicators that self-consistently
point to the explosion occurring not more than $\sim 1,750$ years ago.
These include the current morphology and X-ray luminosity of the Ridge,
the diffuse X-rays from the central $2-3$ parsecs, which apparently
point to stellar winds as the sole contributors to the ISM within
the cavity, and the timing and $2-200$ keV luminosity associated
with the remnant's interaction with the 50 km s$^{-1}$ molecular cloud,
which apparently irradiated Sgr~B2 and, possibly, the ionized halo 
surrounding the Galactic center several hundred years ago. 

On the theoretical plane, several additional steps remain to be taken.
These include a more thorough examination of the nonthermal particle
injection and radiation at the remnant-cloud interface, and a subsequent
analysis of the time-dependent spectrum illuminating Sgr~B2 and other
nearby clouds. Eventually, when the necessary computational resources
become available, it would be very helpful to redo these calculations
in 3 dimensions. Observationally, this work would benefit from a
high resolution mapping of the metal abundances surrounding the explosion
site. The morphology of this distribution should be directly coupled
to the energetics of the explosion and the age of the remnant.

{\bf Acknowledgments} This work was funded in part under the auspices
of the U.S.\ Dept.\ of Energy, and supported by its contract
W-7405-ENG-36 to Los Alamos National Laboratory, by a DOE SciDAC grant
DE-FC02-01ER41176.  At the University of Arizona, this research was
supported by NSF grant AST-0402502, and has made use of
NASA's Astrophysics Data System Abstract Service.  F. M. is grateful
to the University of Melbourne for its support (through a Sir Thomas
Lyle Fellowship and a Miegunyah Fellowship).  The simulations were
conducted on the Space Simulator at Los Alamos National Laboratory.

{}

\begin{deluxetable}{lcccc}
\tablewidth{0pt}
\tablecaption{Simulation Properties\label{table:sims}}
\tablehead{
  \colhead{Simulation}
& \colhead{Energy}
& \colhead{$D_{\rm Sgr A*}^{\rm min}$\tablenotemark{a}}
& \colhead{$T_{\rm GMC}$\tablenotemark{b}}
& \colhead{Remnant Age} \\

& \colhead{($10^{51}$\,erg)}
& \colhead{(pc)}
& \colhead{(y)}
& \colhead{(y)}

}
\startdata

Standard & \phn1.5 & 0.4 & 1,200 & 1,700 \\
Energetic & 12\phd\phn & 0.2 & \phn\phm{,}400 & \phn\phm{,}700 \\

\enddata
\tablenotetext{a}{The distance of closest approach of the shock to
  Sgr~A*}
\tablenotetext{b}{The time it would have taken for the shock to reach
  the 50~km~s$^{-1}$ Giant Molecular Cloud}
\end{deluxetable}
\clearpage

\newpage
\clearpage

\newpage
\clearpage

\newpage
\clearpage

\newpage
\clearpage

\newpage
\clearpage

\newpage
\clearpage

\newpage
\clearpage

\newpage
\clearpage

\newpage
\clearpage

\end{document}